\documentclass[aps,prl,showpacs,twocolumn]{revtex4-1}
\usepackage{amsmath}
\usepackage{amssymb}
\usepackage{graphics}
\usepackage{empheq}
\usepackage{lipsum}

\newcommand\ro{\hat\rho}
\newcommand\Ho{\hat H}
\newcommand\Ao{\hat A}
\newcommand\Bo{\hat B}

\newcommand\qo{\hat q}
\newcommand\po{\hat p}
\newcommand\phio{\hat\phi}

\newcommand\Fo{\hat F}

\newcommand\Mc{\mathcal{M}}

\newcommand\crm{\mathrm{c}}

\newcommand\Tr{\mathrm{Tr}}

\newcommand\half{\frac{1}{2}}

\begin{document}
\title{Exact closed master equation for Gaussian non-Markovian dynamics}   
\author{L. Ferialdi}
\email{ferialdi@math.lmu.de}
\affiliation{Mathematisches Institut, Ludwig-Maximilians Universit\"at, Theresienstr. 39, 80333 Munich.}
\date{\today}
\begin{abstract}
 Non-Markovian master equations describe general open quantum systems when no approximation is made. We provide the exact closed master equation for the class of Gaussian, completely positive, trace preserving, non-Markovian dynamics. This very general result allows to investigate a vast variety of physical systems. We show that the master equation for non-Markovian quantum Brownian motion is a particular case of our general result. Furthermore, we derive the master equation unraveled by a non-Markovian, dissipative stochastic Schr\"odinger equation, paving the way for the analysis of dissipative non-Markovian collapse models.
\end{abstract}
\maketitle

{\it Introduction.}-- 
The theory of open quantum systems strongly developed in the last decades, pushed by technological demand. The basic tools to analyze open quantum systems are Master Equations (MEs), that allow for insight into physical systems by computing average physical quantities. Markovian MEs provide an efficient description of a vast amount of physical processes. These MEs are very well known, their features deeply studied and they obey a precise structure~\cite{BrePet02}. However, in recent years the interest in understanding quantum dynamics beyond the Markov approximation has grown, due to the discovery of many physical systems for which the Markovian description fails~\cite{Leeetal}. Indeed, the timescales of ultra-fast processes are of the same order as the timescale of the bath they interact with, leading to failure of the Markov approximation. Due to the difficulty in treating non-Markovian dynamics, only few exact MEs are known in the literature, most of which are formal, or concerning peculiar systems or stochastic processes~\cite{HPZ,ShaLid05}.

 Since every ME allows for an infinite number of stochastic unravellings, Stochastic Schr\"odinger Equations (SSEs) are an equally powerful tool as MEs. SSEs play an important role in many fields like, e.g., continuos quantum measurement~\cite{Dio98,Dio08}, quantum optics~\cite{Pleetal}, light harvesting systems~\cite{Rodetal}, and foundations of quantum mechanics. In this field, collapse models provide a solution of the measurement problem by describing the evolution of quantum systems by means of SSEs~\cite{BasGhi03}. These models are experiencing renewed interest because they might provide a test of the quantum-to-classical transition in a not so far future~\cite{Bahetal}. It is, in general, extremely difficult to obtain the ME associated to a Gaussian non-Markovian SSE, mainly because the non-local terms displayed by the SSE make it difficult to derive a closed equation.


An important step towards a general understanding of non-Markovian dynamics has recently been taken in~\cite{DioFer14}, where the authors obtained both the most general superoperator and SSE for Gaussian, non-Markovian dynamics. This result is very important because it gives a general characterization of a wide class of non-Markovian dynamics, that can be heuristically obtained from the Lindblad structure. However, the result is rather formal, and it cannot be exploited to compute physical quantities, preventing the explicit analysis of physical systems.
The  aim of this Letter is to fill this gap by deriving a closed ME for a general trace-preserving, completely positive (CP), Gaussian open system dynamics. Moreover, we provide the ME associated to a general Gaussian SSE, both in the dissipative and non-dissipative cases.
We stress that the results obtained are analytical and exact, no approximation has been made.

{\it Derivation of the ME.}-- 
Consider a system bilinearly interacting with a bosonic bath:
\begin{equation}\label{coupling}
\Ao^j(t)\phio_j(t)\,,
\end{equation} 
where $\Ao^j(t)=e^{i\Ho_S t}\Ao^je^{-i\Ho_S t}$ are Hermitian system operators, $\phio_j(t)=e^{i\Ho_B t}\phio_je^{-i\Ho_B t}$ are bosonic fields of the bath, and $H_S$, $H_B$ are respectively system and bath Hamiltonians. The Einstein's sum rule is understood. The generalization to non-Hemitian  $\Ao$ can be easily obtained by expressing them as linear combinations of Hermitian operators. We assume the commutation relations among the systems operators to be:
\begin{equation}\label{comm}
\left[\Ao^j(t),\Ao^k(s)\right]=f^{jk}(t,s)\,,
\end{equation}
with $f^{jk}(t,s)$ an antisymmetric complex functions. 
We assume the initial state of the system to be factorized and that the bath has an initial Gaussian state, fully characterized by the correlation function
\begin{equation}\label{phiphiD}
\Tr_B\left[\phio_j(\tau)\phio_k(s)\ro_B\right]=D_{jk}(\tau,s)\;.
\end{equation}
 We introduce the left-right (LR) formalism \cite{Choetal85,DioFer14}, 
denoting by a subscript $L$ ($R$) the operators acting on $\ro$ from the left (right), 
e.g. $\Ao^k_{L}\Ao^j_{R}\ro=\Ao^k\ro\Ao^j$. We also define the following operators:  $\Ao^j_\Delta=\Ao^j_L-\Ao^j_R$ and $\Ao^j_\crm=(\Ao^j_L+\Ao^j_R)/2$ \cite{Choetal85,DioFer14}. Note that these operators represent respectively a commutator and (half) an anticommutator. It has been recently proved that the system reduced density matrix evolves according to the following equation
\begin{equation}\label{GMt}
\ro_t=\Mc_t\ro_0
\end{equation}
where the most general CP, trace preserving, Gaussian superoperator $\Mc_t$ reads~\cite{DioFer14}
\begin{equation}\label{GMtexp}
\Mc_t=T\exp\left\{-\int_0^t\!d\tau\Ao^j_{\Delta}(\tau)\int_0^{\tau}\!ds\Bo_{j}(\tau,s)
\right\}\,.
\end{equation}
Here $T$ denotes the time ordering operator, and
\begin{equation}\label{Bj}
\Bo_j(\tau,s)= D_{jk}^\mathrm{Re}(\tau,s)\Ao^k_{\Delta}(s)+ 2i D_{jk}^\mathrm{Im}(\tau,s)\Ao^k_{\crm}(s)\,,
\end{equation}
where $D^\mathrm{Re}$ and $D^\mathrm{Im}$ are  respectively real symmetric and imaginary antisymmetric parts of $D$.
Furthermore, in~\cite{DioFer14}  the authors proved that the most general SSE with linear coupling $\Ao^j(t)\phi_j(t)$ that unravels Eq.~\eqref{GMtexp} reads
\begin{equation}\label{SSEdiss}
\frac{d\psi_t}{dt}=-i\Ao^j(t)
                           \!\left(\!\phi_j(t)\!+\!\int_0^t\!ds 
                                 [D_{\!jk}\!(\!t\!,\!s)\!-\!S_{\!jk}\!(\!t\!,\!s)]\frac{\delta}{\delta\phi_k(s)}\!\right)
\!\psi_t\;
\end{equation}
where
\begin{eqnarray}\label{corrnoise}
\mathbb{E}\left[\phi_j^\ast(\tau)\phi_k(s)\right]&=&D_{jk}(\tau,s)\,\\
\mathbb{E}\left[\phi_j(\tau)\phi_k(s)\right]&=&S_{jk}(\tau,s)\,,
\end{eqnarray}
are the correlation functions of the complex, Gaussian, colored noises $\phi_j(t)$.
By setting $D_{jk}(\tau,s)=D_{jk}(\tau)\delta(\tau-s)$ in Eq.~\eqref{GMtexp}, one obtains the Markovian superoperator
\begin{eqnarray}\label{GMtL}
\Mc_t&=&T\exp\left\{\int_0^t\!d\tau D_{jk}(\tau)\right.\\
&&\cdot\!\!\left.\left(\!\Ao^k_{L}(\tau)\Ao^j_{R}(\tau)\!-\!\frac{1}{2}\Ao^j_{L}(\tau)\Ao^k_{L}(\tau)\!-\!\frac{1}{2}\Ao^j_{R}(\tau)\Ao^k_{R}(\tau)\right)\!\!
\right\}.\nonumber
\end{eqnarray}
The presence of the $T$ operator in the map~\eqref{GMtexp} makes it a formal result which cannot be exploited to compute explicitly the evolution of physical quantities. In order to do so one needs to obtain a closed ME. The double integral inside the time ordering operator makes this goal hard to achieve:
one needs to find a way to treat the time ordering of non-local arguments. We expand the map $\Mc_t$~\eqref{GMtexp} in Dyson series:
\begin{equation}\label{Mseries}
\Mc_t=\sum_{n=0}^{\infty} \frac{(-1)^n}{n!}M^n_{t}\,,
\end{equation}
where
\begin{equation}\label{Mn}
M_t^n=T\left[\prod_{i=1}^n\int_0^tdt_i\int_0^{t_i}ds_i\Ao^{j_i}_{\Delta}(t_i)\Bo_{j_i}(t_i,s_i)\right]\,.
\end{equation}
Each $M_t^n$ has two important features that will be used later: they contain the time ordered product (T-product) of $2n$ operators, and they are invariant under  permutation over $i$ of $(t_i, s_i)$. In order to derive the ME, one needs to differentiate Eq.~\eqref{Mn}, obtaining
\begin{equation}\label{MDn}
\dot{M}_t^n\!\!=\!n \Ao^{j_1}_{\Delta}(t) \!\!\!\int_0^{t}\!\!\!ds_1T\!\left[\!\Bo_{j_1}(t,s_1)\!\prod_{i=2}^n\!\int_0^t\!\!\!dt_i\!\!\int_0^{t_i}\!\!\!ds_i\Ao^{j_i}_{\Delta}(t_i)\Bo_{j_i}(t_i,s_i)\!\right]
\end{equation}
where the factor $n$ comes from the afore-mentioned symmetry over $i$ of $M_t^n$.
We stress that while $M_t^n$ contains the T-product of $2n$ operators, $\dot{M}_t^n$ displays $2n-1$ time ordered operators. Since our goal is to express $\dot{\Mc}_t$ in terms of $\Mc_t$, we adopt the following strategy: we write the T-product of an odd number of operators in terms of even T-products. We do so by exploiting Wick's theorem~\cite{Wick}, according to which one can write any T-product as a sum of all the possible contractions of its elements. We denote a Wick contraction with an over-bracket, and we arrange the T-product of Eq.~\eqref{MDn} as follows:
\begin{eqnarray}\label{wick}
\lefteqn{T\!\left[\!\Bo_{j_1}(t,s_1)\!\prod_{i=2}^n\!\int_0^t\!\!\!dt_i\!\!\int_0^{t_i}\!\!\!ds_i\Ao^{j_i}_{\Delta}(t_i)\Bo_{j_i}(t_i,s_i)\right]=}\nonumber\\
&&\Bo_{j_1}(t,s_1) T\!\left[\!\prod_{i=2}^n\!\int_0^t\!\!\!dt_i\!\!\int_0^{t_i}\!\!\!ds_i\Ao^{j_i}_{\Delta}(t_i)\Bo_{j_i}(t_i,s_i)\right]\nonumber\\
&&+\sum_{m=2}^n \int_0^t\!\!\!dt_m\!\!\int_0^{t_m}\!\!\!ds_m\overbracket{\Bo_{j_1}(t,s_1)\Ao^{j_m}_{\Delta}}(t_m)\nonumber\\
&&\,\,\,\,\,\,\,\times T\!\left[\!\Bo_{j_m}(t_m,s_m)\!\prod_{\substack{i=2\\i\neq m}}^n\!\int_0^t\!\!\!dt_i\!\!\int_0^{t_i}\!\!\!ds_i\Ao^{j_i}_{t_i\Delta}\Bo_{j_i}(t_i,s_i)\right]\nonumber\\
&&+\sum_{m=2}^n \int_0^t\!\!\!dt_m\!\!\int_0^{t_m}\!\!\!ds_m\overbracket{\Bo_{j_1}(t,s_1)\Bo_{j_m}(t_m,s_m)}\nonumber\\
&&\,\,\,\,\,\,\,\times T\!\left[\!\Ao^{j_m}_{\Delta}(t_m)\!\prod_{\substack{i=2\\i\neq m}}^n\!\int_0^t\!\!\!dt_i\!\!\int_0^{t_i}\!\!\!ds_i\Ao^{j_i}_{\Delta}(t_i)\Bo_{j_i}(t_i,s_i)\right]
\end{eqnarray}
This equation is nothing but a convenient arrangement of the terms predicted by the Wick's theorem:
the first term of the right hand side accounts for all the contributions not involving contractions of $\Bo(t,s_1)$, while the second and third elements collect the terms involving such contractions. For our problem a contraction is given by
\begin{equation}\label{contr}
\overbracket{\Bo_{j_1}(t,s_1)\Ao^{j_m}_{\Delta}}(t_m)=\left[\Ao^{j_m}_{\Delta}(t_m),\Bo_{j_1}(t,s_1)\right]\theta(t_m-s_1)
\end{equation}
where the unit-step function $\theta$ is needed because we are not using normal ordered products.

Note that the first term of Eq.~\eqref{wick} is a even T-product, while the second and third terms are odd T-products. With the help of Eq.~\eqref{contr}, we decompose these two terms similarly to Eq.~\eqref{wick}, and we iterate this procedure to all the odd T-products obtained. After a long calculation,  exploiting Eq.~\eqref{Mn} one finds that the result of this iteration is
\begin{eqnarray}\label{Tfin}
\lefteqn{T\!\left[\!\Bo_{j_1}(t,s_1)\!\prod_{i=2}^n\!\int_0^t\!\!\!dt_i\!\!\int_0^{t_i}\!\!\!ds_i\Ao^{j_i}_{\Delta}(t_i)\Bo_{j_i}(t_i,s_i)\right]=}\nonumber\\
&&\qquad\qquad\sum_{k=0}^{n-1} c_{j_1}^{n-k-1}(\Ao,\Bo) \frac{(n\!-\!1)!}{k!}M_t^k\qquad\qquad\qquad
\end{eqnarray}
where $c^{n-k-1}_{j_1}(\Ao,\Bo)$ are functionals of $\Ao_{\Delta}$ and $\Bo$, whose analytical expressions are obtained from the recursive substitution just performed. The explicit calculation leading to Eq.~\eqref{Tfin}, as well as the explicit expressions of the $c^n$ are reported in~\cite{sup}. Substituting Eq.~\eqref{Tfin} in Eq.~\eqref{MDn}, after some manipulation one eventually finds that~\cite{sup}:
\begin{eqnarray}\label{MDseries}
\dot{\Mc}_t&=&-\Ao^{j}_{\Delta}(t)\left[\int_0^tds_1\sum_{n=1}^{\infty} (-1)^nc_{j}^{n}(\Ao,\Bo)\right]\Mc_t
\end{eqnarray}
Applying this expression to $\ro_0$, one finds
\begin{equation}\label{MEint}
\dot{\ro}_t\!=\!-\Ao^j_{\Delta}(t)\!\!\left[\int_0^t\!\!\!ds_1\mathcal{A}_{jk}(t,s_1\!)\Ao^k_{\Delta}(s_1\!)\!+\!2i\mathcal{B}_{jk}(t,s_1\!)\Ao^k_{c}(s_1\!)\!\right]\!\ro_t
\end{equation}
where 
\begin{eqnarray}
\label{mathA}\mathcal{A}_{jk}(t,s_1)&=&D^{\mathrm{Re}}_{jk}(t,s_1)+\sum_{n=1}^{\infty}\alpha_{jk}^n(t,s_1)\,\\
\label{mathB}\mathcal{B}_{jk}(t,s_1)&=&D^{\mathrm{Im}}_{jk}(t,s_1)+\sum_{n=1}^{\infty}\beta_{jk}^n(t,s_1)\,,
\end{eqnarray}
and the functions $\alpha^n$, $\beta^n$ are suitable combinations of $D^{\mathrm{Re}}$ and $D^{\mathrm{Im}}$, obtained from the $c^n$~\cite{sup}. We stress that the kernels $\mathcal{A}$, $\mathcal{B}$ are determined analytically.

Equation~\eqref{MEint} is the most general closed ME for a trace preserving, CP, Gaussian, non-Markovian dynamics. It is a time-local ME that, as one expects for a Gaussian dynamics, displays a quadratic dependence on the system operators.

If the system under study is linear, i.e. the free Hamiltonian is at most quadratic, one can further simplify Eq.~\eqref{MEint}. In this case indeed, since the system operators evolve with the free Hamiltonian $\Ho_0$, their Heisenberg equations are linear. Such a system of coupled equations can always be solved univocally by setting two boundary condition. In particular, if we choose as boundary values $\Ao^k(t)$ and $\dot{\Ao}^k(t)$, the solution can be written as follows:
\begin{equation}
\Ao^j(s_1)=\mathcal{C}^{j}_k(t-s_1)\Ao^k(t)+\tilde{\mathcal{C}}^{j}_k(t-s_1)\dot{\Ao}^k(t)
\end{equation}
where $\mathcal{C}$, $\tilde{\mathcal{C}}$ are specific kernels that explicitly depend on the Hamiltonian, and that satisfy $\mathcal{C}(0)=-\dot{\tilde{\mathcal{C}}}(0)=1$ and $\dot{\mathcal{C}}(0)=\tilde{\mathcal{C}}(0)=0$.
Substituting this expression in~\eqref{MEint} one obtains the following time-local closed ME:
\begin{eqnarray}\label{ME}
\dot{\ro}_t&=&\left[\Gamma_{jk}(t) \Ao^j_{\Delta}(t)\Ao^k_{\Delta}(t) + \Theta_{jk}(t) \Ao^j_{\Delta}(t)\dot{\Ao}^k_{\Delta}(t) \right.\nonumber\\
&&\left.+\Xi_{jk}(t)\Ao^j_{\Delta}(t)\Ao^k_{c}(t)+\Upsilon_{jk}(t)\Ao^j_{\Delta}(t)\dot{\Ao}^k_{c}(t)\right]\ro_t\,,
\end{eqnarray}
where
\begin{eqnarray}
\label{Gamma}\Gamma_{jk}(t)&=&-\int_0^tds_1\mathcal{A}_{jl}(t,s_1)\mathcal{C}_{k}^l(t-s_1)\\
\label{Theta}\Theta_{jk}(t)&=&-\int_0^tds_1\mathcal{A}_{jl}(t,s_1)\tilde{\mathcal{C}}_{k}^l(t-s_1)\\
\label{Xi}\Xi_{jk}(t)&=&-2i\int_0^tds_1\mathcal{B}_{jl}(t,s_1)\mathcal{C}_{k}^l(t-s_1)\\
\label{Upsilon}\Upsilon_{jk}(t)&=&-2i\int_0^tds_1\mathcal{B}_{jl}(t,s_1)\tilde{\mathcal{C}}_{k}^l(t-s_1)
\end{eqnarray}
Switching to the Schr\"odinger picture one can write the ME in a more familiar way:
\begin{eqnarray}
\dot{\ro}_t&=&-i[\Ho_0,\ro]+ \Gamma_{jk}(t)[\Ao^j,[\Ao^k,\ro]] +\Theta_{jk}(t) [\Ao^j,[\dot{\Ao}^k,\ro]] \nonumber\\
\label{MEsch}&&+\Xi_{jk}(t)[\Ao^j,\{\Ao^k,\ro\}]+\Upsilon_{jk}(t)[\Ao^j,\{\dot{\Ao}^k,\ro\}]\,.
\end{eqnarray}
Equations~\eqref{ME}-\eqref{MEsch} are the main result of this paper, i.e. the most general trace preserving, CP, Gaussian, non-Markovian ME for a linear system. This ME is the generalization of the Lindblad ME to non-Markovian dynamics. We stress that this result is exact and all the functions entering these equations are analytical. We observe that the ME is characterized by an explicit dependence not only on $\Ao(t)$, but also on $\dot{\Ao}(t)$. This is a purely non-Markovian feature, since the Lindblad superoperator~\eqref{GMtL} instead depends on $\Ao(t)$ only. 
More comments on this issue are given later with specific examples.

We now focus on some interesting physical systems, and we exploit our general achievement to provide the ME unraveled by a non-Markovian, dissipative SSE.

{\it Non-Markovian quantum Brownian motion}.-- In their seminal paper~\cite{HPZ}, Hu, Paz and Zhang derived the ME for a particle interacting with an environment of harmonic oscillators using the path-integral formalism. Since the Hamiltonian they analyzed is a particular case of our general result, we provide a straightforward way to give an alternative derivation of such a ME. The system considered in~\cite{HPZ} is an harmonic oscillator of mass $m$ coupled to the bath via the position operator. The ME is easily obtained substituting $\Ao=\qo$ in Eq.~\eqref{MEsch} (only one $\Ao$):
\begin{eqnarray}\label{HPZ}
\dot{\ro}_t&=&-i[\Ho_0,\ro]+ \Gamma(t)[\qo,[\qo,\ro]] +m\Theta(t) [\qo,[\po,\ro]] \nonumber\\
&&+\Xi(t)[\qo^2,\ro]+m\Upsilon(t)[\qo,\{\po,\ro\}]\,.
\end{eqnarray}
This is the same ME obtained by Hu-Paz-Zhang, it is CP and it describes quantum dissipation in the non-Markovian regime. The functions $\Gamma$, $\Theta$, $\Xi$, $\Upsilon$  are given by Eqs.~\eqref{Gamma}-\eqref{Upsilon} with $\mathcal{C}(t-s_1)=\cos\omega(t-s_1)$ and $\tilde{\mathcal{C}}(t-s_1)=-\sin\omega(t-s_1)/m\omega$, and they display a series structure coming from $\mathcal{A}$ and $\mathcal{B}$. 
The equivalence of the functions $\Gamma$, $\Theta$, $\Xi$, $\Upsilon$ with those obtained by Hu-Paz-Zhang can be easily checked in the weak coupling limit, i.e. considering only the \lq\lq zero order\rq\rq~terms of Eqs.~\eqref{mathA}-\eqref{mathB} (cf. Eqs.(2.46a)-(2.46d) of~\cite{HPZ}). 
Nonetheless, since the coefficients of Eq.~\eqref{HPZ} have a series structure, our derivation provides a straightforward way to obtain higher orders expansions of such coefficients. 

Caldeira and Leggett derived a Markov limit of the CP ME~\eqref{HPZ}, obtaining a non CP ME (Eq.~\eqref{HPZ} with $\Theta=0$)~\cite{CalLeg83}. 
The common explanation of this fact is that a term of the type $[\po,[\po,\ro]]$ is missing because the limiting procedure is such that it is lost. 
What we argue is that CP is broken by the dissipative term $[\qo,\{\po,\ro\}]$.
 Indeed, such a term is a contribution of the  $\dot{\Ao}$ type ($\po\sim\dot{\qo}$) that, as previously stressed, is not expected in a Markovian ME. Such an unexpected term arises from the limiting procedure, and we believe that this is the real issue with it. Accordingly, in the Markovian regime one cannot correctly describe dissipation by considering a system-bath coupling only via $\qo$.
As we will clarify with the next example, the correct way to describe Markovian quantum dissipation is by considering a system-bath coupling involving both $\qo$ and $\po$ (as already suggested in~\cite{Dio93}).

{\it Non-Markvovian dissipative SSE.}-- Another interesting application of our main result is the derivation of the ME associated to a non-Markovian dissipative SSE. Our starting point is a given SSE and we derive the ME which is unraveled by it. We consider the QMUPL collapse model, that is particularly interesting because it offers itself for a detailed mathematical analysis~\cite{Dio90,Basetal,Feretal}. The ME for the non-Markovian QMUPL model has never been computed, nor in the dissipative and non-dissipative cases: this result will allow to analyze the physical features of this model that could not be investigated so far. The SSE describing the non-Markovian dissipative QMUPL model reads~\cite{Feretal}:
\begin{eqnarray}\label{NMdissSE}
\frac{d}{dt}|\psi_t\rangle&=&\left[-\frac{i}{\hbar}\left(\Ho_0+\frac{\lambda\mu}{2}\{\qo,\po\}\right)+\sqrt{\lambda}\left(\qo+i\frac{\mu}{\hbar}\po\right)\phi(t)\right.\nonumber\\
&&\left.-2\sqrt{\lambda}\qo\int_0^t ds\, D(t,s)\frac{\delta}{\delta \phi(s)}\right]|\psi_t\rangle\,.
\end{eqnarray}
The integral term and the fact that the real noise $\phi$ is coupled both to $\qo$ and $\po$, make it hard to derive a closed  ME with standard techniques (e.g. path integration). This equation can be rewritten in the form~\eqref{SSEdiss} by defining the following operators $\Ao_1=\hbar\qo$, $\Ao_2=-\mu\po$, and noises $\phi_1=i\sqrt{\lambda}\phi$, $\phi_2=\sqrt{\lambda}\phi$. Given these prescriptions one can immediately obtain the ME for this model from Eq.~\eqref{MEsch}. In particular if $\Ho_0$ is an harmonic oscillator one obtains the following result:
\begin{eqnarray}\label{NMdissME}
\dot{\ro}_t&=&-i[\Ho(t),\ro]+ \Gamma(t)[\qo,[\qo,\ro]] +\Theta(t) [\qo,[\po,\ro]] \nonumber\\
&&+\Xi(t)[\qo^2,\ro]+\Upsilon(t)[\qo,\{\po,\ro\}]+\gamma(t)[\po,[\po,\ro]]\,,
\end{eqnarray}
where $\Ho(t)=\Ho_0+\alpha(t)\po^2+(\beta(t)+\frac{\lambda\mu}{2})\{\qo,\po\}$, and the explicit expressions of the coefficients are given in~\cite{sup}.
This ME displays all the terms entering the ME describing non-Markovian Brownian motion, plus three further contributions. Two of them ($[\po^2,\ro]$ and $[\{\qo,\po\},\ro]$) are responsible for energy renormalization, while the term $[\po,[\po,\ro]]$ is a new contribution to diffusion. The existence of these new terms is due to the fact that in Eq.~\eqref{NMdissSE} the interaction is mediated both by the position and momentum operators, while the open system leading to Eq.~\eqref{HPZ} interacts with the bath only via $\qo$. Indeed, one can easily check that the ME~\eqref{HPZ} is unraveled by the SSE~\eqref{NMdissSE} when the coupling $\qo+i\po$ is replaced with $\qo+i\qo$. As expected, the white noise limit of Eq.~\eqref{NMdissME} recovers the ME obtained in~\cite{BasIppVac05} for the Markovian dissipative QMUPL model. This Markovian ME, has the same structure of the one describing CP Quantum (Markovian) Brownian motion~\cite{Dio93,Dio95}. 

We stress that if one considers an open system coupled to a bath both via $\qo$ and $\po$ (and not only $\qo$, like in the Hu-Paz-Zhang and Caldeira-Leggett MEs), its reduced dynamics is described by the ME~\eqref{NMdissME}. As already mentioned, this ME leads to the correct CP dissipative ME under the Markov limit. The dissipative term containing $\po$ is now a legitimate Markovian contribution ($\Ao$ type) because $\po$ enters the coupling. Moreover, also the term $[\po,[\po,\ro]]$ naturally emerges from the $\po$-coupling and does not need to be added by hand. These facts suggest that the correct way to describe dissipation in the Markov regime is by considering a system-bath coupling mediated both by $\qo$ and $\po$, as it is implicitly assumed in collisional models~\cite{Dio95}. The physical intuition is the following: dissipation is a dynamical feature, and as such it requires a dynamical description. In Markovian dynamics, the instantaneous interaction between the system and the bath erases any dynamical effect of the bath on the system, and to keep track of dissipation one needs a dynamical coupling (i.e. $\po$). In the non-Markovian regime instead, the dynamics has memory of the interaction, and the $\qo$ coupling is sufficient to describe dissipation (e.g. the Hu-Paz-Zhang ME). Of course, if one wants to coherently describe dissipation both in the Markovian and non-Markovian regimes (allowing for a smooth transition between them), one should consider the coupling via $\qo$ and $\po$ also in the non-Markovian case.

{\it Non-Markovian non-dissipative dynamics.}--  An open system dynamics is non-dissipative when $D^{\mathrm{Im}}=0$. Applying this restriction to Eqs.~\eqref{mathA}-\eqref{mathB} one finds that the kernels $\Xi_{jk}(t)$, $\Upsilon_{jk}(t)$ do not contribute to the ME. The only contribution comes from the parts of $\Gamma_{jk}(t)$ and $\Theta_{jk}(t)$ proportional to $D^\mathrm{Re}$. Accordingly, the ME~\eqref{MEsch} in the non-dissipative case reads:
\begin{equation}\label{MEnodiss}
\dot{\ro}_t=-i[\Ho_0,\ro]+ \tilde{\Gamma}_{jk}(t)[\Ao^j,[\Ao^k,\ro]]+\tilde{\Theta}_{jk}(t)[\Ao^j,[\dot{\Ao}^k,\ro]] 
\end{equation}
with
\begin{eqnarray}
\label{Gammatil}\tilde{\Gamma}_{jk}(t)&=&-\int_0^tds_1D_{jl}^{\mathrm{Re}}(t,s_1)\mathcal{C}_{k}^l(t-s_1)\\
\label{Thetatil}\tilde{\Theta}_{jk}(t)&=&-\int_0^tds_1D_{jl}^{\mathrm{Re}}(t,s_1)\tilde{\mathcal{C}}_{k}^l(t-s_1)
\end{eqnarray}
If the system under study is an harmonic oscillator with proper frequency $\omega$, the prescription $\Ao=\sqrt{\lambda}\qo$ in Eq.~\eqref{MEnodiss} leads to
\begin{equation}\label{NM-QMUPL}
\dot{\ro}_t=-i[\Ho,\ro]+ \lambda\tilde{\Gamma}(t)[\qo,[\qo,\ro]] +\lambda\tilde{\Theta}(t) [\qo,[\po,\ro]]\,,
\end{equation}
where  $\tilde{\Gamma}$ and  $\tilde{\Theta}$ are given by Eqs.~\eqref{Gammatil}-\eqref{Thetatil} with $\mathcal{C}(t-s)= \cos \omega(t-s)$ and $\tilde{\mathcal{C}}(t-s)= - \sin \omega(t-s)/m\omega$.
This ME is the non-Markovian generalization of the Joos-Zeh ME~\cite{JoosZeh}, and displays a purely non-Markovian contribution to diffusion ($[\qo,[\po,\ro]]$). The white noise limit ($D^{\mathrm{Re}}(t,s)\rightarrow\delta(t-s)$) of Eqs.~\eqref{Gammatil}-\eqref{Thetatil} gives respectively $\Gamma\rightarrow-1$ and $\Theta\rightarrow0$, leading Eq.~\eqref{NM-QMUPL} to recover the ME by Joos-Zeh.  

We stress that, if one considers the momentum coupling, the requirement of non-dissipative dynamics coincides with setting $\mu=0$. One can easily check that by imposing this requirement on Eq.~\eqref{NMdissME} one recovers Eq.~\eqref{NM-QMUPL}. Moreover, the SSE unraveling Eq.~\eqref{NM-QMUPL} is obtained by setting $\mu=0$ in Eq.~\eqref{NMdissSE}. This is the SSE describing the non-dissipative, non-Markovian QMUPL model~\cite{Basetal}.

{\it Conclusions.}-- 
In this Letter we provide the exact closed ME for a wide class of non-Markovian dynamics, both for linear and non-linear systems. This very general result allows to investigate many physical systems and, under suitable conditions, it recovers the ME for non-Markovian quantum Brownian motion. The derivation of the ME for a system coupled to a bath both via $\qo$ and $\po$, suggests that quantum dissipation is better described by a model of this kind (instead of the $\qo$ coupling only). We showed that our general result provides a straightforward tool to obtain the ME  unraveled by a non-Markovian SSE, allowing for a deeper analysis of the physical features of non-Markovian collapse models. These features, as well as the analysis of the ME~\eqref{NMdissME} for an open system will be investigated in a forthcoming paper.
The procedure outlined in this Letter can also be applied to the pure dephasing spin-boson model, the reason being that the free Hamiltonian commutes with the interaction Hamiltonian and the commutation relation~\eqref{comm} is satisfied. Substituting $\Ao=\sigma_z$ in Eq.~\eqref{MEsch} one recovers the known ME for this model~\cite{BrePet02,GuaSmiVac12}. 
The general spin-boson model can be analyzed only by generalizing the method here proposed. 
This topic will be subject of further studies. 
The achievement of this Letter will improve the understanding of non-Markovianity under the physical point of view.

\phantom{}
The author was supported by the Marie Curie Fellowship PIEF-GA-2012-328600. The author wishes to thank A. Smirne for many useful conversations and valuable comments on a first draft of the manuscript.

\newpage

\renewcommand{\theequation}{S.\arabic{equation}}
\setcounter{equation}{0}
\widetext{
\begin{center}
{\bf \large Supplementary material}
\end{center}

\vspace{0.5cm}
{\bf Derivation of the master equation (18).} The calculations leading to the main result of our paper are rather involved. In this Supplementary Material we provide the technical details of the derivation of Eq.~(17).

We start from Eq.(14) of the main text, which can be simplified by exploiting the symmetry of $M_n$ over permutation of the time labels $(t_i,s_i)$. Namely, the terms entering the sums are all equal, only the labeling changes. Accordingly, one can rewrite the sum over the contractions of Eq.(14) as $n-1$ times one specific contraction, leading to:
\begin{eqnarray}\label{wick2}
\lefteqn{T\!\left[\!\Bo_{j_1}(t,s_1)\!\prod_{i=2}^n\!\int_0^t\!\!\!dt_i\!\!\int_0^{t_i}\!\!\!ds_i\Ao^{j_i}_{\Delta}(t_i)\Bo_{j_i}(t_i,s_i)\right]=\Bo_{j_1}(t,s_1) T\!\left[\!\prod_{i=2}^n\!\int_0^t\!\!\!dt_i\!\!\int_0^{t_i}\!\!\!ds_i\Ao^{j_i}_{\Delta}(t_i)\Bo_{j_i}(t_i,s_i)\right]}\\
&&+(n-1)\int_0^t\!\!\!dt_2\!\!\int_0^{t_2}\!\!\!ds_2\overbracket{\Bo_{j_1}(t,s_1)\Ao^{j_2}_{\Delta}(t_2)}T\!\left[\!\Bo_{j_2}(t_2,s_2)\!\prod_{i=3}^n\!\int_0^t\!\!\!dt_i\!\!\int_0^{t_i}\!\!\!ds_i\Ao^{j_i}_{\Delta}(t_i)\Bo^{j_i}(t_i,s_i)\right]\nonumber\\
&&+(n-1) \int_0^t\!\!\!dt_2\!\!\int_0^{t_2}\!\!\!ds_2\overbracket{\Bo_{j_1}(t,s_1)\Bo_{j_2}(t_2,s_2)}T\!\left[\!\Ao^{j_2}_{\Delta}(t_2)\!\prod_{i=3}^n\!\int_0^t\!\!\!dt_i\!\!\int_0^{t_i}\!\!\!ds_i\Ao^{j_i}_{\Delta}(t_i)\Bo_{j_i}(t_i,s_i)\right]\,,\nonumber
\end{eqnarray}
where
\begin{eqnarray}\label{contr}
\overbracket{\Bo_{j_1}(t,s_1)\Ao^{j_2}_{\Delta}(t_2)}&=& 2i D_{j_1k}^{\mathrm{Im}}(t,s_1)f^{j_2k}(t_2,s_1) \theta(t_2-s_1)\,,\\
\overbracket{\Bo_{j_1}(t,s_1)\Bo_{j_2}(t_2,s_2)}&=& 2i  \left[D_{j_1k}^{\mathrm{Im}}(t,s_1)D_{j_2l}^{\mathrm{Re}}(t_2,s_2)+D_{j_1k}^{\mathrm{Re}}(t,s_1)D_{j_2l}^{\mathrm{Im}}(t_2,s_2)\right]f^{lk}(s_1,s_2)\theta(s_2-s_1)\,.
\end{eqnarray}
We observe that the first term of the right hand side of Eq.~\eqref{wick2} is of the type we want, since it is an even T-product. The second and third terms instead display an odd number of operators to be time ordered and they need to be rewritten. One then applies Eq.~(14) to the second (third) term of Eq.~\eqref{wick2} obtaining, similarly to Eq.~\eqref{wick2}, an even T-products and two series containing contractions of $\Bo$ ($\Ao$).
Applying this procedure recursively, one eventually obtains a series expansion of the initial odd T-product, exclusively in terms of even T-products of lower order:
\begin{eqnarray}
T\!\left[\!\Bo_{j_1}(t,s_1)\!\prod_{i=2}^n\!\int_0^t\!\!\!dt_i\!\!\int_0^{t_i}\!\!\!ds_i\Ao^{j_i}_{\Delta}(t_i)\Bo_{j_i}(t_i,s_i)\right]&=&\sum_{k=0}^{n-1} c_{j_1}^{n-k-1}(\Ao,\Bo) \frac{(n-1)!}{k!}T\!\left[\!\prod_{i=1}^k\!\int_0^t\!\!\!dt_i\!\!\int_0^{t_i}\!\!\!ds_i\Ao^{j_i}_{\Delta}(t_i)\Bo_{j_i}(t_i,s_i)\right]\nonumber\\
\label{Tfin}&=&\sum_{k=0}^{n-1} c_{j_1}^{n-k-1}(\Ao,\Bo) \frac{(n-1)!}{k!}M_t^k\,,
\end{eqnarray}
where Eq.~\eqref{Tfin} is obtained by exploiting Eq.~(12).
Substituting this result in Eq. (13), and exploiting Eq.~(11) one can write $\dot{\Mc}_t$ as follows:
\begin{equation}\label{MDseries}
\dot{\Mc}_t=
-\Ao^{j_1}_{\Delta}(t)\int_0^tds_1\sum_{n=1}^{\infty}(-1)^{n-1}\sum_{k=0}^{n-1}  \frac{c_{j_1}^{n-k-1}(\Ao,\Bo)}{k!}M^k_t\,.
\end{equation}
Performing the change of label $m=n-1$, and exploiting the definition of Cauchy product of two series,  we achieve the goal of expressing $\dot{\Mc}_t$ in terms of $\Mc_t$:
\begin{equation}
\dot{\Mc}_t=-\Ao^j_{\Delta}(t)\left[\int_0^tds_1\sum_{n=0}^{\infty} (-1)^n c^{n}_j(\Ao,\Bo)\right]\Mc_t\,.
\end{equation}
The $c^{n}_j(\Ao,\Bo)$ are functionals of $\Ao_{\Delta}$ and $\Bo$, and their explicit expressions are involved. A long calculation leads to $c_j^0=\Bo_j(t,s_1)$, and
\begin{equation}\label{coeffs}
c_j^{n}(\Ao,\Bo)= \int_0^tdt_2\int_0^{t_2}ds_2\left(b_j^{n,j_2}(s_1,t_2)\Bo_{j_2}(t_2,s_2)+a^n_{jj_2}(s_1,s_2)\Ao^{j_2}_{\Delta}(t_2)\right)\,,
\end{equation}
where $n\geq 1$, and the comma on the superscript distinguishes the series index $n$ from the dummy index $j_2$. The $b^n$ and $a^n$ are recursively defined as follows
\begin{eqnarray}
b_j^{1,j_2}(s_1,t_2)&=&\overbracket{\Bo_j(t,s_1)\Ao^{j_2}_{\Delta}(t_2)}\,,\\
b_j^{n,j_2}(s_1,t_2)&=&\int_0^tdt_{n+1}\int_0^{t_{n+1}}ds_{n+1}b_j^{1,k}(s_1,t_{n+1})b_k^{n-1,j_2}(s_{n+1},t_2)\,,\quad n\geq2\,,\\
a_{jj_2}^1(s_1,s_2)&=&\overbracket{\Bo_{j}(t,s_1)\Bo_{j_2}(t_2,s_2)}\,,\\
a_{jj_2}^n(s_1,s_2)&=&\int_0^t\!\!\!dt_{n+1}\int_0^{t_{n+1}}\!\!\!\!ds_{n+1}a_{jk}^1(s_1,t_{n+1})a_{kj_2}^{n-1}(s_{n+1},s_2)+a^{n-1}_{jk}(s_1,s_{n+1})\overbracket{\Ao^k_{\Delta}(t_{n+1})\Bo_{j_2}(t_2,s_2)}\,,\,\, n\geq2.
\end{eqnarray}
Note that, since both $b^1(s_1,t_2)$ and $a^1(s_1,s_2)$  depend on $D^\mathrm{Im}$, by construction also every $b^n(s_1,t_2)$ and $a^n(s_1,s_2)$ display such dependence. This fact plays a crucial role when considering non-dissipative dynamics, i.e.  $D^\mathrm{Im}=0$.
At this stage we make use of the definition of $\Bo$ of Eq.(6), in order to make explicit the dependence on $\Ao_{\Delta}(s_1)$ and $\Ao_{c}(s_1)$. The result is
\begin{equation}\label{coeffs2}
\sum_{n=0}^{\infty} (-1)^n c^{n}(\Ao,\Bo)=\mathcal{A}_{jk}(s_1) \Ao^k_{\Delta}(s_1)+2i\mathcal{B}_{jk}(s_1) \Ao^k_{c}(s_1)\,,
\end{equation}
where
\begin{eqnarray}
\label{mathA}\mathcal{A}_{jk}(t,s_1)&=& D^\mathrm{Re}_{jk}(t,s_1)+\sum_{n=1}^{\infty}\alpha_{jk}^n(t,s_1)\,,\\
\label{mathB}\mathcal{B}_{jk}(t,s_1)&=& D^\mathrm{Im}_{jk}(t,s_1)+\sum_{n=1}^{\infty}\beta_{jk}^n(t,s_1)\,,
\end{eqnarray}
and
\begin{eqnarray}
\alpha^n_{jk}(t,s_1)&=&(-1)^n \left(\int_{s_1}^tds_2\int_0^tdt_2b^{n,l}_j(t_2,s_2)D^\mathrm{Re}_{lk}(s_2,s_1)+\int_0^{s_1}ds_2\int_0^tdt_2a^n_{jk}(t_2,s_2)\right)\,,\\
\beta^n_{jk}(t,s_1)&=& (-1)^n\int_{s_1}^tds_2\int_0^tdt_2b^{n,l}_j(t_2,s_2)D^\mathrm{Im}_{lk}(s_2,s_1)\,.
\end{eqnarray}
Exploiting these definitions one can determine analytically both the functions of the ME (18) and those displayed by the ME (27) for linear systems.
Furthermore, we mention that ME (27) can be recast in the non diagonal Kossakovski form as follows:
\begin{equation}
\label{MEnodiag}\dot{\ro}_t=-i[\Ho,\ro]+\!\!\!\sum_{l,m=1,2}\!\!\! f_{jk}^{lm}(t)\!\left(\!\Fo_l^j\ro F_m^k-\half\{F_m^kF_l^j,\ro\}\!\right).
\end{equation}
where $\Ho=\Ho_0+i\Xi_{jk}(t) \Ao^j\Ao^k+i\Upsilon_{jk}(t)\{\Ao^j,\dot{\Ao}^k\}$, $\Fo_1^j=\Ao^j$, $\Fo_2^j=\dot{\Ao}^j$, and
\begin{equation}\label{f}
f_{jk}^{lm}(t)=\left(
\begin{array}{cc}
\Gamma_{jk}(t)&\Theta_{jk}(t)+\Upsilon_{jk}(t)\\
\Theta_{jk}(t)-\Upsilon_{jk}(t)&0
\end{array}\right)
\end{equation}

\vspace{0.5cm}
{\bf Explicit expressions of the coefficients of Eq.~(32).} The system under study is an harmonic oscillator of mass $m$ and proper frequency $\omega$:
\begin{equation}
\Ho=\frac{\po^2}{2m}+\frac{m\omega^2}{2}\qo^2+\frac{\lambda\mu}{2}\{\qo,\po\}
\end{equation}
The Heisenberg equations of motion are linear and their solution reads:
\begin{equation}
\left(\begin{array}{c}
\qo_s\\
\po_s
\end{array}\right)=\mathcal{C}(t-s)
\left(\begin{array}{c}
\qo_t\\
\po_t
\end{array}\right)
\end{equation}
with
\begin{equation}
\mathcal{C}(t-s)=
\left(\begin{array}{cc}
\cos\tilde{\omega}(t-s)-\frac{\lambda\mu}{\tilde{\omega}}\sin\tilde{\omega}(t-s)&-\frac{1}{m\tilde{\omega}}\sin\tilde{\omega}(t-s)\\
\frac{m\omega^2}{\tilde{\omega}}\sin\tilde{\omega}(t-s)&\cos\tilde{\omega}(t-s)+\frac{\lambda\mu}{\tilde{\omega}}\sin\tilde{\omega}(t-s)
\end{array}\right)\,,
\end{equation}
where we have defined $\tilde{\omega}=\sqrt{\omega^2-\lambda^2\mu^2}$. Note that since $\po$ is in the coupling, $\tilde{\mathcal{C}}_{ij}(t-s)=\mathcal{C}_{ji}(t-s)$. Substituting these expressions in Eqs.~(23)-(26) of the Letter, one finds:
\begin{eqnarray}
\label{Gamma}\Gamma(t)&=&-\int_0^t\mathcal{A}_{11}(t,s)\mathcal{C}_{11}(t-s)-\mu\mathcal{A}_{12}(t,s)\mathcal{C}_{21}(t-s)ds\\
\label{Theta}\Theta(t)&=&-\int_0^t\mathcal{A}_{11}(t,s)\mathcal{C}_{12}(t-s)-\mu\mathcal{A}_{12}(t,s)\mathcal{C}_{22}(t-s)-\mu\mathcal{A}_{21}(t,s)\mathcal{C}_{11}(t-s)+\mu^2\mathcal{A}_{22}(t,s)\mathcal{C}_{21}(t-s)ds\\
\label{Xi}\Xi(t)&=&-i\int_0^t\mathcal{B}_{11}(t,s)\mathcal{C}_{11}(t-s)-\mu\mathcal{B}_{12}(t,s)\mathcal{C}_{21}(t-s)ds\\
\label{Upsilon}\Upsilon(t)&=&-i\int_0^t\mathcal{B}_{11}(t,s)\mathcal{C}_{12}(t-s)-\mu\mathcal{B}_{12}(t,s)\mathcal{C}_{22}(t-s)-\mu\mathcal{B}_{21}(t,s)\mathcal{C}_{11}(t-s)+\mu^2\mathcal{B}_{22}(t,s)\mathcal{C}_{21}(t-s)ds\\
\alpha(t)&=&\int_0^ti\mu\mathcal{B}_{21}(t,s)\mathcal{C}_{12}(t-s)-i\mu^2\mathcal{B}_{22}(t,s)\mathcal{C}_{22}(t-s)ds\\
\beta(t)&=&\int_0^ti\mu\mathcal{B}_{21}(t,s)\mathcal{C}_{11}(t-s)-i\mu^2\mathcal{B}_{22}(t,s)\mathcal{C}_{21}(t-s)ds\\
\gamma(t)&=&\int_0^t\mu\mathcal{A}_{21}(t,s)\mathcal{C}_{12}(t-s)-\mu^2\mathcal{A}_{22}(t,s)\mathcal{C}_{22}(t-s)ds 
\end{eqnarray}
where $\mathcal{A}$ and $\mathcal{B}$ are given by Eqs.~\eqref{mathA}-\eqref{mathB} with the prescriptions
\begin{equation}
D^{\mathrm{Re}}_{11}(t,s)=-D^{\mathrm{Im}}_{12}(t,s)=D^{\mathrm{Im}}_{21}(t,s)=D^{\mathrm{Re}}_{22}(t,s)\equiv\lambda D(t,s)
\end{equation}
\begin{equation}
D^{\mathrm{Im}}_{11}(t,s)=D^{\mathrm{Re}}_{12}(t,s)=D^{\mathrm{Re}}_{21}(t,s)=D^{\mathrm{Im}}_{22}(t,s)\equiv0
\end{equation}
}


\begin{thebibliography}{99}
\bibitem{BrePet02} H.P. Breuer and F. Petruccione Theory of open quantum systems (Oxford, Oxford University Press, 2002).

\bibitem{Leeetal}    L. S. Cederbaum, E. Gindensperger, I. Burghardt, Phys. Rev. Lett. {\bf 94}, 113003 (2005); H. Lee, Y-C. Cheng, G. R. Fleming, Science {\bf 316}, 1462 (2007); G. D. Scholes {\it et al.}, Nature Chem. {\bf 3}, 763 (2011); P.Rebentrost, A. Aspuru-Guzik, J. Chem. Phys. {\bf 134}, 101103 (2011); T. Gu\'erin {\it et al.} Nat. Chem. {\bf 4}, 568 (2012); S. Gršblacher, A. Trubarov, N. Prigge, G. D. Cole, M. Aspelmeyer, J. Eisert, Nature Communications {\bf 6}, 7606 (2015).

\bibitem{HPZ} B.L. Hu, J. P. Paz, Y. Zhang, Phys Rev. D {\bf 45}, 2843 (1992).

\bibitem{ShaLid05} A. Shabani, D. A. Lidar, Phys. Rev. A {\bf 71}, 020101(R) (2005); M. W. Y. Tu and W. M. Zhang, Phys. Rev. B {\bf 78}, 235311 (2008); H. P. Breuer, B. Vacchini, Phys. Rev. E {\bf 79}, 041147 (2009); W.-M. Zhang, P.-Y. Lo, H.-N. Xiong, M. W.-Y. Tu, F. Nori, Phys. Rev. Lett.  {\bf 109}, 170402 (2012); B. Vacchini, Phys. Rev. A {\bf 87}, 030101(R) (2013).

\bibitem{Dio98} L. Di\'osi, N. Gisin and W.T. Strunz,  Phys. Rev. A {\bf 58}, 1699 (1998); W.T. Strunz, L. Di\'osi, and N. Gisin, Phys. Rev. Lett. {\bf 82}, 1801 (1999); J. T. Stockburger, H. Grabert, Phys. Rev Lett. {\bf 88}, 170407 (2002).

\bibitem{Dio08} L. Di\'osi, Phys. Rev. Lett. {\bf 100}, 080401 (2008); H.M. Wiseman and J.M. Gambetta, Phys. Rev. Lett. {\bf 101}, 140401 (2008); L. Di\'osi, Phys. Rev. Lett. {\bf 101}, 149902(E) (2008).

\bibitem{Pleetal} M.B. Plenio, P.L. Knight, Rev. Mod. Phys. {\bf 70}, 101 (1998); T. Yu, Phys. Rev A {\bf 69}, 062107 (2004).

\bibitem{Rodetal} J. Roden, W. T. Strunz, A. Eisfeld, J. Chem. Phys. {\bf 134}, 034902 (2011); G. Ritschel, J. Roden, W. T. Strunz, A. Aspuru-Guzik, A. Eisfeld,  J. Phys. Chem. Lett. {\bf 2}, 2912 (2011);
G. Ritschel, D. Suess, S. M\"obius, W. T. Strunz, A. Eisfeld, J. Chem. Phys. {\bf 142}, 034115 (2015).

\bibitem{BasGhi03} A.Bassi and G.C.Ghirardi, Phys. Rep. {\bf 379}, 257 (2003).

\bibitem{Bahetal} A. Bassi, K. Lochan, S. Satin, T. P. Singh, H. Ulbricht, Rev. Mod. Phys. {\bf 85}, 471 (2013); M. Bahrami, M. Paternostro, A. Bassi, H. Ulbricht, Phys. Rev. Lett. {\bf 112}, 210404 (2014).

\bibitem{DioFer14} L. Di\'osi, L. Ferialdi, Phys. Rev. Lett. {\bf 113}, 200403 (2014).

\bibitem{Choetal85} K. Chou, Z. Su, B. Hao and L. Yu, Found. Rep. {\bf 118}, 1 (1985); L. Di\'osi, Found. Phys. {\bf 20}, 63 (1990).

\bibitem{Wick} G. C. Wick, Phys. Rev. {\bf 80}, 268 (1950).

\bibitem{sup} See Supplemental Material at [URL will be inserted by publisher] for mathematical details of the results of this Letter.

\bibitem{CalLeg83} A.O. Caldeira, A. Leggett, Physica A {\bf 121}, 587 (1983).

\bibitem{Dio93} L. Diosi, Europhys. Lett. {\bf 22}, 1 (1993); J. Halliwell, A. Zoupas, Phys. Rev. D {\bf 52}, 7294 (1995); S. Gao, Phys. Rev. Lett. {\bf 79}, 3101 (1997); G. W. Ford, R. F. O'Connell, Phys. Rev. Lett. {\bf 82}, 3376 (1999).

\bibitem{Dio90} L. Di\'osi, Phys. Rev. A {\bf 40}, 1165 (1990).

\bibitem{Basetal}  A. Bassi and L. Ferialdi, Phys. Rev. A {\bf 80}, 012116 (2009); Phys. Rev. Lett. {\bf 103}, 050403 (2009).

\bibitem{Feretal} L. Ferialdi and A. Bassi Phys. Rev. Lett. {\bf 108}, 170404 (2012); Phys. Rev. A {\bf 86}, 022108 (2012).


\bibitem{BasIppVac05} A. Bassi, E. Ippoliti, B. Vacchini,  J. Phys. A {\bf 38}, 8017 (2005).

\bibitem{Dio95} L. Diosi, Europhys. Lett. {\bf 30}, 63 (1995); B. Vacchini, Phys. Rev Lett. {\bf 84}, 1374 (2000).

\bibitem{JoosZeh} E. Joos, H. D. Zeh, Z. Phys. B {\bf 59}, 223 (1985); E. Joos, et al., {\it Decoherence and the Appearance of a Classical World in Quantum Theory}, Springer (2003).

\bibitem{GuaSmiVac12} G. Guarnieri, A. Smirne, B. Vacchini, Phys. Rev. A {\bf 90}, 022110 (2014).

\end{thebibliography}
\end{document}